\documentclass[submission,copyright,creativecommons]{eptcs}
\providecommand{\event}{SR 2013} % Name of the event you are submitting to

\usepackage{amssymb}
\usepackage[english]{babel}
\usepackage[utf8]{inputenc}
\usepackage[babel]{csquotes}

\usepackage{amsmath}
\usepackage{stmaryrd}% For \llbracket, \rrbracket, etc.
\usepackage{mathtools}% For \overbracket, \underbracket.
\usepackage{amsthm}
\usepackage{enumitem}% For choosing enumeration styles.
\usepackage{listings}
\usepackage{tikz}
\usepackage{caption}% For figures.
\usepackage{subcaption}% For subfigures.
\usepackage{calc}% Needed for the definition of the \phantomas command below.
\usepackage{underscore}% Solves problem with underscores in BibTex URLs, see http://handy-it-commands.blogspot.de/2012/01/solving-bibtex-problem-with-underscores.html
\usepackage{framed}

\usetikzlibrary{automata}
\usetikzlibrary{backgrounds}% For framed tikz pictures.
\usetikzlibrary{shapes}% For cloud shapes.
\usetikzlibrary{patterns}
\tikzstyle{every picture}+=[shorten >=1pt, node distance=2cm, inner sep=1mm, auto]
\tikzstyle{initial}+=[initial text=]
\tikzstyle{every state}+=[minimum size=5mm]

\DeclareSymbolFont{extra}{OML}{cmm}{m}{it}
\DeclareMathSymbol{\varrho}{\mathord}{extra}{'045}

\newtheorem{theorem}{Theorem}
\tikzstyle{every picture}+=[node distance=2cm, inner sep=1mm, auto]
\tikzstyle{initial}+=[initial text=]
\tikzstyle{every state}+=[minimum size=8mm]
\tikzstyle{active}+=[fill=red!40]
\tikzstyle{channel_cell}+=[rectangle, draw, minimum size=7mm]
\tikzstyle{marked}+=[draw=blue, line width=1]

\lstdefinelanguage{reactivelanguage}{
	morekeywords={while,input,output,skip,if,then,else,;,assign,call},
	morekeywords=[2]{True,False,iff,xor},
	sensitive=false,
	morecomment=[l]{//},
	morecomment=[s]{/*}{*/},
	morestring=[b]",
}

\lstdefinelanguage{pseudolanguage}{
	morekeywords={while,input,output,skip,if,then,else,assign,call,
	              repeat,until,do,return,end},
	morekeywords=[2]{True,False,iff,xor},
	sensitive=false,
	morecomment=[l]{//},
	morecomment=[s]{/*}{*/},
	morestring=[b]",
	alsoletter={chr(32)},
}

\newcommand*\phantomas[3][c]{%
   \ifmmode
     \makebox[\widthof{$#2$}][#1]{$#3$}%
   \else
     \makebox[\widthof{#2}][#1]{#3}%
   \fi
}

\newcommand{\overbar}[1]{\mkern 1.5mu\overline{\mkern-1.5mu#1\mkern-1.5mu}\mkern 1.5mu}% An overline/bar compromise: just the "right" length.

\newcommand{\ignore}[1]{}
\newcommand{\z}[1]{\text{#1}}

\newcommand{\set}[1]{\left\{ #1 \right\}}
\newcommand{\condset}[2]{\left\{ \, #1 \, \mid \, #2 \, \right\}}

\newcommand{\aut}[1]{\mathcal{#1}}
\newcommand{\specaut}{\aut{A}_{\overbar{R}}}
\newcommand{\lang}[1]{\mathcal{#1}}

\newcommand{\f}[1]{\mathit{#1}}
\newcommand{\sembrack}[1]{\llbracket #1 \rrbracket}
\newcommand{\sembrackk}[1]{\langle\langle #1 \rangle\rangle}%{\llparenthesis #1 \rrparenthesis}

\newcommand{\squeezetrans}[1]{#1}%{\smash{\raisebox{-0.35ex}{$\scriptstyle#1$}}}
\newcommand{\transt}[2][]{\xrightarrow{\squeezetrans{#2}}_{#1}}

\newcommand{\pcode}[1]{\textup{\tt{#1}}}% \ŧexttt will not work here for some reason ...

\newcommand{\cocfgk}{\f{CoCfg}_k(B,\specaut)}

\newcommand{\kfg}{\gamma}

\newcommand{\sigmain}{\mathbb{B}}
\newcommand{\sigmaout}{\mathbb{B}}
\title{Synthesizing Structured Reactive Programs \\ via Deterministic Tree Automata}
\author{Benedikt Brütsch
	\institute{RWTH Aachen University, Lehrstuhl für Informatik 7, Germany}
	\email{bruetsch@automata.rwth-aachen.de}
}
\def\titlerunning{Synthesizing Structured Reactive Programs via Deterministic Tree Automata}
\def\authorrunning{Benedikt Brütsch}

\AtBeginDocument{
	\hypersetup{
		pdftitle = {\titlerunning},
		pdfauthor = {\authorrunning},
		pdfkeywords = {synthesis, reactive programs, tree automata}
	}
}

\begin{document}

\maketitle

\begin{abstract}
Existing approaches to the synthesis of reactive systems typically involve the
construction of transition systems such as Mealy automata.
However, in order to obtain a succinct representation of the desired system,
structured programs can be a more suitable model.
In 2011, Madhusudan proposed an algorithm to construct a structured
reactive program for a given $\omega$-regular specification without
synthesizing a transition system first.
His procedure is based on two-way alternating $\omega$-automata on finite
trees that recognize the set of "correct" programs.

We present a more elementary and direct approach using only deterministic
bottom-up tree automata that compute so-called \emph{signatures}
for a given program.
In doing so, we extend Madhusudan's results to the wider class
of programs with bounded delay, which may read several input symbols before
producing an output symbol (or vice versa).
As a formal foundation, we inductively define a semantics for such programs.
\end{abstract}

\vspace{2mm}
\section{Introduction}

Algorithmic synthesis is a rapidly developing field with many application
areas such as reactive sytems, planning and economics.
% The synthesis problem for reactive systems was originally formulated by Church
% \cite{church1957applications}.
% In a game-theoretic context, solving this problem amounts to finding a
% winning strategy for a controller in an infinite two-player game against the
% environment.
Most approaches to the synthesis of reactive systems, for instance
\cite{buchi_solving_1969,Rabin:1972:AIO:540412,DBLP:conf/popl/PnueliR89,kupferman_churchs_1999},
revolve around synthesizing transition systems such as Mealy or Moore
automata.
%This applies to the first solutions to Church's problem by Büchi and Landweber
%\cite{buchi_solving_1969}, and Rabin \cite{Rabin:1972:AIO:540412}, but also
%to more recent approaches \cite{PnueliR89,Schewe,TODO}.
Unfortunately, the resulting transition systems can be very large.
This has motivated the development of techniques for the reduction of their
state space (for example, \cite{DBLP:journals/corr/abs-1102-4120}).
Furthermore, the method of bounded synthesis
\cite{schewe_bounded_2007,ehlers_symbolic_2010} can be used to synthesize
minimal transition systems by iteratively increasing the bound on the size
of the resulting system until a solution is found.\ignore{TODO: Is this an appropriate description?}
However, it is not always possible to obtain small transition systems.
For example, for certain specifications in linear temporal logic
(LTL), the size of the smallest transition systems satisfying these
specifications is doubly exponential in the length of the formula
\cite{rosner}.

Aminof, Mogavero and Murano \cite{aminof_synthesis_2012} provide a
round-based algorithm to synthesize hierarchical transition systems, which can be
exponentially more succinct than corresponding "flat" transition systems.
The desired system is constructed in a bottom-up manner:
In each round, a specification is provided and the algorithm constructs a
corresponding hierarchical transition system from a given library of available
components and the hierarchical
transition systems created in previous rounds.
Thus, in order to obtain a small system in the last
round, the specifications in the previous rounds have to be chosen in an
appropriate way.

Current techniques for the synthesis of (potentially) succinct implementations in the form
of circuits or programs typically proceed in an
indirect way, by converting a transition system into such an implementation.
For example, Bloem et al.\ \cite{Bloem20073}
first construct a symbolic representation (a binary decision diagram) of an
appropriate transition system and then extract a corresponding circuit.
However, this indirect approach does not necessarily yield a succinct result.

Madhusudan addresses this issue in \cite{madhusudan:LIPIcs:2011:3247}, where
he proposes a procedure to synthesize programs
without computing a transition system first. He considers
\emph{structured reactive programs} over a given set of Boolean variables,
which can be significantly smaller (regarding the length of the
program code) than equivalent transition systems.
To some degree, these programs separate control flow
from memory.
Such a separation can also be found in a related approach that has recently
been introduced by Gelderie \cite{Gelderie12}, where strategies for
infinite games are represented by \emph{strategy machines}, which are
equipped with control states and a memory tape.

Given a finite set of Boolean variables and a nondeterministic Büchi automaton
recognizing the complement of the specification,
Madhusudan constructs a two-way
alternating $\omega$-automaton on finite trees that recognizes the set of
\emph{all} programs over these variables that satisfy the specification.
This automaton can be transformed into a nondeterministic tree automaton (NTA)
to check for emptiness and extract a minimal program
(regarding the height of the corresponding tree) from that set.
In contrast to the transition systems constructed by classical synthesis
algorithms, the synthesized program does not depend on the specific syntactic
formulation of the specification, but only on its meaning.

In this paper, we present a direct construction of a deterministic bottom-up
tree automaton (DTA) recognizing the set of correct programs, without a detour
via more intricate types of automata.
The DTA inductively computes a representation of the behavior of a given
program in the form of so-called \emph{signatures}.
A similar representation is used by Lustig and Vardi in their work on the
synthesis of reactive systems from component libraries
\cite{Lustig09} to characterize the behavior of\ignore{TODO: Adjective here?}
the components.

Our approach is not limited to programs that read input and write output in
strict alternation, but extends Madhusudan's results to the more general class
of programs with \emph{bounded delay}:
In general, a program may read multiple input symbols before writing the next
output symbol, or vice versa, causing a delay between the input sequence and
the output sequence.
In a game-theoretic context, such a program corresponds to a strategy for a
controller in a game against the environment where in each move
the controller is allowed to either choose at least one output symbol or
skip and wait for the next input (see \cite{Holtmann10}).
\ignore{Reading multiple input symbols in a row is necessary to satisfy specifications
that require a certain amount of lookahead on the input sequence.
On the other hand, writing multiple output symbols in a row may
reduce\ignore{TODO! Other word?} the required number of program variables,
because once an output symbol has been written, no information about that
symbol has to be stored anymore.}
%because it can reduce the
%need to store future output symbols that are already determined by the
%previous input sequence.
We consider programs that never cause a delay greater than a given bound
$k \in \mathbb{N}$.

\ignore{
Note that reading multiple input symbols in a row can be necessary if the
specification requires a certain amount of lookahead on the input sequence.
On the other hand, writing multiple output symbols before reading the next
input symbol may reduce the required number of program variables.
For example, consider the specification that requires for some fixed
$m \in \mathbb{N}$ that the $(i+m)$th output symbol agrees with the $i$th
input symbol, for all $i \in \mathbb{N}$.
Any program that reads input and writes output in strict alternation needs at
least $m+1$ Boolean variables to satisfy the specification.
However, if the first $m$ output symbols are written
before reading the first input symbol, a single variable suffices.
}

For a fixed $k$, the complexity of our construction matches that of
Madhusudan's algorithm.
In particular, the size of the resulting DTA is exponential in the size of the
given nondeterministic Büchi automaton recognizing the complement of the
specification, and doubly exponential in the number of program variables.
In fact, we establish a lower bound, showing that the set of all programs
over $n$ Boolean variables that satisfy a given specification cannot even be
recognized by an NTA with less than $2^{2^{n-1}}$ states,
if any such programs exist.
However, note that a DTA (or NTA) accepting precisely these programs enables
us to extract a minimal program for the given specification and the given set
of program variables.
Hence, the synthesized program itself might be rather small.

To lay a foundation for our study of the synthesis of structured reactive
programs, we define a formal semantics for such programs, which is only
informally indicated by Madhusudan.
To that end, we introduce the concept of
\emph{Input/Output/Internal machines (IOI machines)},
which are composable in the same way as structured programs.
This allows for an inductive definition of the semantics.
\ignore{TODO: Point to technical report?}

\vspace{5mm}
\section{Syntax and Semantics of Structured Programs}

We consider a slight modification of the structured
programming language defined in \cite{madhusudan:LIPIcs:2011:3247},
using only single Boolean values as input and output symbols to simplify
notation.
\emph{Expressions} and \emph{programs} over a finite set $B$ of
Boolean variables
are defined by the following grammar, where $b \in B$:
\vspace{6pt}
\newcommand{\tbar}{\enspace\,|\,\enspace}%
\newcommand{\ph}[1]{\langle\mathit{#1}\rangle}
\begin{align*}
	\begin{array}{ccl}
		\ph{expr} & \Coloneqq & \pcode{true} \tbar \pcode{false} \tbar b \tbar
								\ph{expr}\land\ph{expr} \tbar \ph{expr}\lor\ph{expr} \tbar \neg\ph{expr}
	\end{array}
	\\[2pt]
	\begin{array}{ccl}
		\ph{prog} & \Coloneqq & b \coloneqq \ph{expr} \tbar \pcode{input }b \tbar \pcode{output }b \tbar
								\ph{prog}\pcode{;}\ph{prog} \\
				&			&	\pcode{if } \ph{expr} \pcode{ then } \ph{prog} \pcode{ else } \ph{prog} \tbar
								\pcode{while } \ph{expr} \pcode{ do } \ph{prog}
	\end{array}
\end{align*}

%\vspace{-4pt}
Intuitively, ``$\pcode{input } b$'' reads a Boolean value and
stores it in the variable $b$.
Conversely, ``$\pcode{output } b$'' writes the current value of $b$.
To define a formal semantics we associate
with each program a so-called
\emph{IOI machine}.
An IOI machine is a transition system with designated entry and exit
states.
It can have input, output and internal
transitions, with labels of the form $(a_\z{in},\varepsilon)$,
$(\varepsilon,a_\z{out})$ or $(\varepsilon,\varepsilon)$, respectively,
where $a_\z{in}, a_\z{out} \in \mathbb{B}=\set{0,1}$.
An IOI machine is equipped with a finite set $B$ of Boolean variables,
whose valuation is uniquely determined at each state.
A \emph{valuation}
is a function $\sigma\colon B \to \mathbb{B}$ that assigns a Boolean value
to each variable.

The associated IOI machine of an atomic program (i.e., an input statement,
output statement or assignment) has one entry state and exit state for
each possible variable valuation,
and its transitions lead from entry states to
exit states.
\ignore{Figure \ref{atomicioi} shows the associated IOI machine for the atomic program
``$\pcode{input } b$'' over $B = \set{b}$.
\begin{figure}[h!]\label{atomicioi}
\centering
\begin{subfigure}[h!]{9cm}
\centering	
\begin{framed}
\begin{tikzpicture}
	\tikzstyle{every state}+=[rounded rectangle]
	\node[state] (p0) at (0,0) {$b=0$};
	\node[state] (p1) at (0,-2) {$b=1$};
	\node[state,rectangle] (q0) at (6,0) {$b=0$};
	\node[state,rectangle] (q1) at (6	,-2) {$b=1$};
	
	\path[->] (p0) edge node {$(0,\varepsilon)$} (q0);
	\path[->] (p0) edge node[sloped,above=3mm,left=4mm] {$(1,\varepsilon)$} (q1);
	\path[->] (p1) edge node[sloped,above=3mm,left=12mm] {$(0,\varepsilon)$} (q0);
	\path[->] (p1) edge node {$(1,\varepsilon)$} (q1);
\end{tikzpicture}
\end{framed}
\end{subfigure}
\caption{The associated IOI machine of ``$\pcode{input } b$'' for $B=\set{b}$.
	Entry states have rounded corners, exit states have non-rounded corners.\ignore{TODO!}}
\end{figure}
}
For example, at each entry state of the associated IOI machine of an atomic
program of the form ``$\pcode{input } b$'', there are two outgoing input transitions
-- one for each possible input symbol.
The target of such an input transition is the exit state whose variable
valuation is obtained by replacing the value of $b$ with the respective
input symbol.\ignore{TODO: Better explanation?}
The IOI machine of a composite program can be constructed
inductively from the IOI machines of its subprograms,
leveraging their entry and exit states and the
variable valuations of these states.

A \emph{computation} $\varrho$ of a program is a finite or
infinite sequence of subsequent transitions of the corresponding IOI machine:
%\vspace{-8pt}\ignore{TODO: Check the spacing!}
\[
	\varrho = q_1 \transt{(a_1,b_1)} q_2 \transt{(a_2,b_2)} q_3 \transt{(a_3,b_3)} \cdots
\]
The \emph{label} of $\varrho$ is the pair of finite or
infinite words
$(a_1 a_2 a_3 \ldots, \; b_1 b_2 b_3 \ldots) \in
	(\sigmain^* \cup \sigmain^\omega) \times (\sigmaout^* \cup \sigmaout^\omega)$.
An \emph{initial computation} starts at the unique entry
state where all variables have the value $0$.
The \emph{infinite behavior} $\sembrackk{p}$ of a program $p$
is the set of infinite input/output sequences
$(\alpha,\beta) \in \mathbb{B}^\omega \times \mathbb{B}^\omega$
that can be produced by an initial computation of $p$.
Furthermore, we call a program \emph{reactive} if all its initial computations
can be extended to infinite computations that yield an infinite input
and output sequence.

At any given time during a computation $\varrho$ as above, the length of the
input sequence $a_1a_2 \ldots a_i$ and the output sequence $b_1b_2 \ldots b_i$
might differ.
The supremum of these length differences along a computation is called the
\emph{delay} of the computation.
If the delay of a computation does not exceed a given bound
$k \in \mathbb{N}$ then we call this computation \emph{$k$-bounded}.
A program is said to be $k$-bounded if all its computations are $k$-bounded.
By restricting the infinite behavior of a program $p$ to labels of
$k$-bounded initial computations,
we obtain the \emph{$k$-bounded infinite behavior} $\sembrackk{p}_k$ of $p$.

\section{Solving the Synthesis Problem Using Deterministic Tree Automata}

The synthesis problem for structured reactive programs with bounded delay
can be formulated as follows:
Given an $\omega$-regular specification
$R \subseteq \left(\sigmain \times \sigmaout\right)^\omega$ representing
the permissible input/output sequences,
a finite set of Boolean variables $B$ and a delay bound $k \in \mathbb{N}$,
the task is to construct a structured reactive program $p$ over $B$ with
$k$-bounded delay such that $\sembrackk{p} \subseteq R$ -- or detect that
no such program exists.
(However, our results can easily be generalized to finite input and
output alphabets other than $\mathbb{B}$ by allowing input and output
statements that process multiple Boolean values as in
\cite{madhusudan:LIPIcs:2011:3247}.)
In the following we assume that the specification $R$ is provided in the
form of a \emph{nondeterministic Büchi automaton (NBA)} $\specaut$
over the alphabet $\sigmain \times \sigmaout$ that recognizes the
complement of the specification, i.e.,
$\lang{L}(\specaut) =
	\left(\sigmain \times \sigmaout\right)^\omega \setminus R$,
which is always possible for $\omega$-regular specifications.

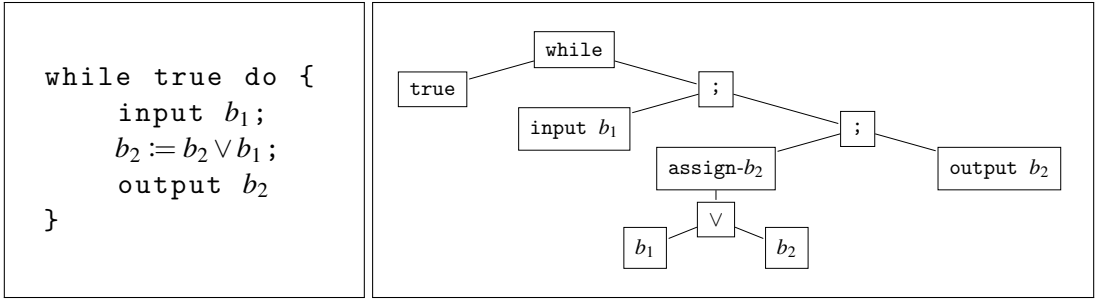
\begin{figure}[h!]
\centering

\setlength{\FrameSep}{\fboxsep-0.5mm}% TODO: This is a hack to decrease space inside the frame.
\begin{subfigure}[h]{0.3\linewidth}\ignore{TODO: Better example?}
%\Fbox{
\begin{framed}
\begin{minipage}[c][3.8cm][c]{\linewidth}% TODO: The lines in the code start with blanks to adjust spacing inside the frame. This is a hack.
\begin{lstlisting}[linewidth=\dimexpr\linewidth-8mm\relax,xleftmargin=2mm]
 while true do {
     input $b_1$;
     $b_2 \coloneqq b_2 \lor b_1$;
     output $b_2$
 }
\end{lstlisting}
\end{minipage}
\end{framed}
%}
\end{subfigure}
\begin{subfigure}[h]{0.6\linewidth}
%\fbox{
\begin{framed}
\begin{minipage}[c][3.8cm]{\linewidth}
\centering
\vspace{2mm}
\resizebox{\dimexpr\linewidth-5mm\relax}{!}{
%\resizebox{!}{\textheight}{
\begin{tikzpicture}
\tikzstyle{every node}+=[inner sep=2mm,rectangle,draw]
\tikzstyle{level}+=[sibling distance=50mm,level distance=7mm]
\tikzstyle{level 4}+=[level distance=9mm]
\tikzstyle{level 5}+=[sibling distance=25mm,level distance=5mm]

%\node [rectangle,draw] (root){root} [edge from parent fork down]
    %child {
    \node (while) {\texttt{while}}
        child {node (true) {\texttt{true}}}
        child {node (semi1) {\texttt{;}}
            child {node (input) {\texttt{input }$b_1$}}
            child {node (semi2) {\texttt{;}}
                child {node (assign) {\texttt{assign}-$b_2$}
                    child {node (and) {$\lor$}
                        child {node (b1) {$b_1$}}
                        child {node (b2) {$b_2$}}
                    }
                }
                child {node [rectangle,draw] (output) {\texttt{output }$b_2$}}
            }
        };
    %};
\end{tikzpicture}
}
\vspace{2mm}
\end{minipage}
\end{framed}
%}
\end{subfigure}

\caption{Example: A program and its tree representation.}
\label{progtreeex}
\end{figure}

Our synthesis procedure is based on the fact that programs can be
viewed as trees.
Figure \ref{progtreeex} shows an example for a tree representation of a
program.
We use \emph{deterministic bottom-up tree automata} (\emph{DTAs},
see, for example, \cite{rozenberg_languages_1997})
to recognize sets of programs.
More specifically, we show the following theorem:
\begin{theorem}\label{dtatheorem}
	Let $B$ be a finite set of Boolean variables,
	let $k \in \mathbb{N}$ and
	let $\specaut$ be a nondeterministic Büchi automaton recognizing the
	complement of a specification
	$R \subseteq \left(\sigmain \times \sigmaout\right)^\omega$.
	We can construct a DTA that accepts
	a tree $p$ iff $p$ is a reactive program over $B$ with $k$-bounded delay
	and $\sembrackk{p} \subseteq R$, such that the size of this DTA
	is doubly exponential in $|B|$ and $k$ and exponential in the size of
	$\specaut$.
\end{theorem}\ignore{TODO: Time complexity?}

An emptiness test on this DTA yields a solution to the synthesis problem.
We obtain the desired tree automaton by intersecting three DTAs:
The first DTA $\aut{B}_\z{sat}(B,k,\specaut)$ recognizes
the set of programs over $B$ whose $k$-bounded computations satisfy the
specification $R$. That means, a program $p$ is accepted iff
$\sembrackk{p}_k \subseteq R$.
The second DTA $\aut{B}_\z{reactive}(B)$ recognizes the reactive
programs over $B$.
Finally, we use a third DTA $\aut{B}_\z{delay}(B,k)$ to recognize the programs
over $B$ with $k$-bounded delay.
We only consider the construction of $\aut{B}_\z{sat}(B,k,\specaut)$ here,
as the other two DTAs can be constructed in a very similar way.

% TODO: Section?
The DTA $\aut{B}_\z{sat}(B,k,\specaut)$ evaluates a given program $p$ in a
bottom-up manner, thereby assigning one of its states to each node of the
program tree.
The state reached at the root node must provide
enough information to decide whether $\sembrackk{p}_k \subseteq R$,
or equivalently, whether
$\sembrackk{p}_k \cap \lang{L}(\specaut) = \emptyset$.
To that end, we are interested in the possible runs of $\specaut$
on the input/output sequences generated by the program.
Thus, we consider pairs of program computations and corresponding runs of
$\specaut$, which we call \emph{co-executions}.
Intuitively, $\aut{B}_\z{sat}(B,k,\specaut)$ inductively computes a
representation of the possible co-executions of a given program and
$\specaut$.
We define these representations, called \emph{co-execution signatures},
in the following.

The beginning and end of a co-execution can be indicated by
a valuation of the program variables and a state of $\specaut$.
However, we have to consider the following:
The input sequence of a computation might be longer or shorter than
its output sequence, but a run of $\specaut$ only consumes input and
output sequences of the same length.
The suffix of the input/output sequence after the end of the shorter sequence,
called the \emph{overhanging suffix}, is hence still waiting to be
consumed by $\specaut$.
Thus, we indicate the start and end of a co-execution by tuples of the
form $\kfg = (\sigma,s,u,v)$, called \emph{co-configurations},
where $\sigma$ is a variable valuation, $s$ is a state of $\specaut$ and
$(u,v) \in \left(\sigmain^* \times \set{\varepsilon}\right)
	\cup \left(\set{\varepsilon} \times \sigmaout^*\right)$ is an
overhanging suffix.
Since we are only interested in $k$-bounded computations, we only
consider co-configurations with $|u| \leq k$ and $|v| \leq k$.
The set of these co-configurations for a given set of variables $B$ and a
given NBA $\specaut$ is denoted by $\cocfgk$.

A finite co-execution is called \emph{complete} if the program terminates at
the end of the computation.
The \emph{finite co-execution signature} $\f{cosig}^\z{fin}(p,\specaut,k)$
of a program $p$ (with respect to $\specaut$) is a relation consisting of
tuples of the form $(\kfg,f,\kfg\kern+.11em')$ with $f \in \mathbb{B}$,
which indicate that there exists a complete $k$-bounded co-execution that\ignore{TODO: k-bounded co-execution? Definition?}
starts with the co-configuration $\kfg$ and ends with $\kfg\kern+.11em'$ such that the corresponding
run of $\specaut$ visits a final state iff $f=1$.
The \emph{infinite co-execution signature}
$\f{cosig}^\infty(p,\specaut,k)$ of $p$ is a set of co-configurations with
$\gamma \in \f{cosig}^\infty(p,\specaut,k)$ iff there exists an infinite
$k$-bounded co-execution starting with $\kfg$ such that the run of $\specaut$
visits a final state infinitely often.
We use pairs consisting of a finite and infinite co-execution signature as
states of the DTA $\aut{B}_\z{sat}(B,k,\specaut)$.
The size of the DTA is hence determined by the number of possible
co-execution signatures, which is doubly exponential in the number of
variables and $k$ and exponential in the size of $\specaut$.
For a fixed $k$, this matches the complexity of Madhusudan's construction
\cite{madhusudan:LIPIcs:2011:3247}.

If $\sigma_0$ is the initial variable valuation (where all variables have the value
$0$) and $s_0$ is the initial state of $\specaut$, then
$(\sigma_0,s_0,\varepsilon,\varepsilon) \in \f{cosig}^\infty(p,\specaut,k)$
iff there is an initial $k$-bounded computation of $p$ such that some
corresponding run of $\specaut$ visits a final state infinitely often,
so $\f{cosig}^\infty(p,\specaut,k)$ is indeed sufficient to decide whether
$\sembrackk{p}_k \subseteq R$.
It remains to be shown that the co-execution signatures can be computed
inductively.
Exemplarily, we consider the case of programs of the form
$p = \z{``}\texttt{while } e \texttt{ do } p_1\z{''}$.
First, we construct a representation $\f{cosig}^*_e(p_1,\specaut,k)$
of all finite sequences of consecutive
co-executions of $p_1$ that are compatible with the loop condition $e$.
To that end, we consider only
those tuples $(\kfg,f,\kfg\kern+.11em')$ in $\f{cosig}^\z{fin}(p_1,\specaut,k)$
where the variable valuation in $\kfg$
satisfies the loop condition $e$, and compute the reflexive transitive closure
of the resulting relation.
Formally, we have
$\f{cosig}^*_e(p_1,\specaut,k) = \f{closure}(C)$ with
$C = \condset {
			( (\sigma,s,u,v), f, \kfg\kern+.11em' ) \in \f{cosig}^\z{fin}(p_1,\specaut,k)
		}{ \sigma \in \sembrack{e} }$.
Here, $\sembrack{e}$ denotes the set of variable valuations that satisfy $e$,
and $\f{closure}(C)$ is the smallest relation
$D \subseteq \cocfgk \times \mathbb{B} \times \cocfgk$
\,such that
\begin{itemize}
	\item $(\kfg, 0, \kfg) \in D$ for all $\kfg \in \cocfgk$, and
	\item $(\kfg, f_1, \kfg\kern+.11em') \in D, \, (\kfg\kern+.11em', f_2, \kfg\kern+.11em'') \in C$ implies $(\kfg, \max\set{f_1,f_2}, \kfg\kern+.11em'') \in D$.
\end{itemize}

Using $\f{cosig}^*_e(p_1,\specaut,k)$,
the co-execution signatures for $p$ can be
computed by the following reasoning:
A finite co-execution of
$p = \z{``}\texttt{while } e \texttt{ do } p_1\z{''}$ (and $\specaut$)
can be decomposed into a finite sequence of co-executions of $p_1$.
An infinite co-execution of $p$ can either eventually stay
inside a loop iteration forever or traverse infinitely many iterations.
It can therefore be decomposed either into a finite sequence of
co-executions of $p_1$ followed by an infinite co-execution
of $p_1$,
or into a finite sequence of co-executions of $p_1$ followed by
a cycle of co-executions of $p_1$, leading back to a previous
co-configuration.
Thus, we obtain the following formal representation of the co-execution
signatures for $p$:

\begin{itemize}
	\item $( \kfg, f, (\sigma',s',u',v') ) \in \f{cosig}^\z{fin}(p,\specaut,k)$ \; iff \;
		$( \kfg, f, (\sigma',s',u',v') ) \in \f{cosig}^*_e(p_1,\specaut,k)$
		and $\sigma' \notin \sembrack{e}$.
	\item $\kfg \in \f{cosig}^\infty(p,\specaut,k)$ iff at least one of the following holds:
		\begin{itemize}
			\item There exist $\kfg\kern+.11em' = (\sigma',s',u',v') \in \cocfgk$ and
				$f \in \mathbb{B}$ \\
				such that
				$(\kfg, f, \kfg\kern+.11em') \in \f{cosig}^*_e(p_1,\specaut,k)$, \,
				$\sigma' \in \sembrack{e}$ and
				$\kfg\kern+.11em' \in \f{cosig}^\infty(p_1,\specaut,k)$.
			\item There exist $\kfg\kern+.11em' = (\sigma',s',u',v') \in \cocfgk$ and
				$f \in \mathbb{B}$ \\
				such that
				$(\kfg, f, \kfg\kern+.11em') \in \f{cosig}^*_e(p_1,\specaut,k)$, \,
				$\sigma' \in \sembrack{e}$ and
				$(\kfg\kern+.11em', 1, \kfg\kern+.11em') \in \f{cosig}^*_e(p_1,\specaut,k)$.
		\end{itemize}
\end{itemize}

\section{Lower Bound for the Size of the Tree Automata}\label{optimality}

We show the following lower bound for the size of any nondeterministic tree
automaton (NTA) recognizing the desired set of programs:
\begin{theorem}\label{lowerboundnta}
	Let $B$ be a set of $n$ Boolean variables,
	let $k \in \mathbb{N}$ and
	let $R \subseteq \left(\sigmain \times \sigmaout\right)^\omega$
	be a specification that is realizable by some program over $B$ with
	$k$-bounded delay.
	Let $\aut{C}$ be an NTA that accepts
	a tree $p$ iff $p$ is a reactive program over $B$ with $k$-bounded delay
	and $\sembrackk{p} \subseteq R$.
	Then $\aut{C}$ has at least $2^{2^{n-1}}$ states.
\end{theorem}

For a sketch of the proof,
consider a set of Boolean variables $B = \set{b_1,\dotsc,b_n}$.
There are $2^{2^{n-1}}$ functions of the type
$\mathbb{B}^{n-1} \to \mathbb{B}$.
Each of these functions can be implemented by a program that checks the values
of $b_1,\dotsc,b_{n-1}$ and sets $b_n$ to the corresponding function value.
An NTA as in Theorem \ref{lowerboundnta} must be able to distinguish
all of these programs.
Otherwise, let $p_i$ and $p_j$ be two such programs that cannot be
distinguished by the NTA.
We could then construct a program that satisfies the specification and
contains $p_i$ as a subprogram, but runs into a non-reactive infinite loop
if this subprogram is replaced by $p_j$.
The NTA would accept both variants, including the non-reactive program,
which contradicts the premise.

\section{Conclusion}

The contributions of this paper are threefold,
advancing the study of structured reactive programs:
We introduced a formal semantics for structured reactive programs in the sense
of \cite{madhusudan:LIPIcs:2011:3247}.
Furthermore, we presented a new synthesis algorithm for structured reactive
programs with bounded delay, using the elementary concept of deterministic
bottom-up tree automata.
Finally, we showed a lower bound for the size of any nondeterministic tree
automaton that recognizes the set of specification-compliant programs,
emphasizing the importance of choosing a small yet still sufficient set of
program variables.
Estimating the number of Boolean variables that are needed to realize a given
specification is a major open problem.
While \cite{rosner} implies an exponential upper bound for the required
number of variables in the case of LTL specifications, a corresponding lower
bound is still to be determined.

\ignore{
We presented a new synthesis algorithm for structured reactive programs with
bounded delay using the elementary concept of deterministic bottom-up tree
automata.
Our approach yields a characterization of the behavior of a program in the
form of signatures, which can be deduced from the signatures of the
subprograms.
We showed a lower bound for the size of any nondeterministic tree automaton
that recognizes the set of specification-compliant programs,
emphasizing the importance of choosing a small yet still sufficient set of
program variables.
A major open problem is the estimation of the number of Boolean variables
that are needed to realize a given specification.
While \cite{rosner} implies an exponential upper bound for the required
number of variables in the case of LTL specifications, a corresponding lower
bound is still to be determined.
}

\paragraph*{Acknowledgments.}
The author would like to thank Wolfgang Thomas for his helpful advice and
Marcus Gelderie\ignore{TODO: Insert.} for fruitful discussions.

\bibliographystyle{eptcs}
\bibliography{bibliography}

\end{document}